\documentclass[paper,11pt]{article}
\pdfoutput=1

\usepackage[pdftex,bookmarks,colorlinks]{hyperref}
\usepackage[pdftex]{graphicx}
\usepackage[amssymb]{SIunits}
\usepackage{mathrsfs,mathcomp}
\usepackage{fancyhdr}
\fancypagestyle{plain}{
\fancyhead[CO]{\footnotesize{Candidate video for the 30th Annual Gallery of Fluid Motion\\ 65rd Annual Meeting of the American Physical Society, Division of Fluid Dynamics \\San Diego, CA, 18-20 Nov 2012}}
}

\begin{document}

\title{Taming Acoustic Cavitation}

\author{David Fernandez Rivas\footnote{dfrivascu@gmail.com}, Bram Verhaagen, Oscar R. Enr\'iquez,\\ Michel Versluis, Andrea Prosperetti, Han Gardeniers, Detlef Lohse\\
\emph{MESA+ Institute for Nanotechnology, University of Twente}\\
\emph{Enschede, The Netherlands}\\
}

\maketitle

\begin{abstract}

In this fluid dynamics video we show acoustic cavitation occurring from pits etched on a silicon 
surface. By immersing the surface in a liquid, gas pockets are entrapped in the pits which upon ultrasonic insonation, are observed to shed cavitation bubbles. Modulating the driving pressure it is possible to induce different behaviours based on the force balance that determines the interaction among bubbles and the silicon surface.
This system can be used for several applications like sonochemical water treatment, cleaning of surfaces with deposited materials such as biofilms \cite{Fdez1}.

\end{abstract}

\section{Technical Details of the submitted video}

\begin{figure}[h]
\centering
\label{fig1}
\includegraphics[width=0.7\textwidth]{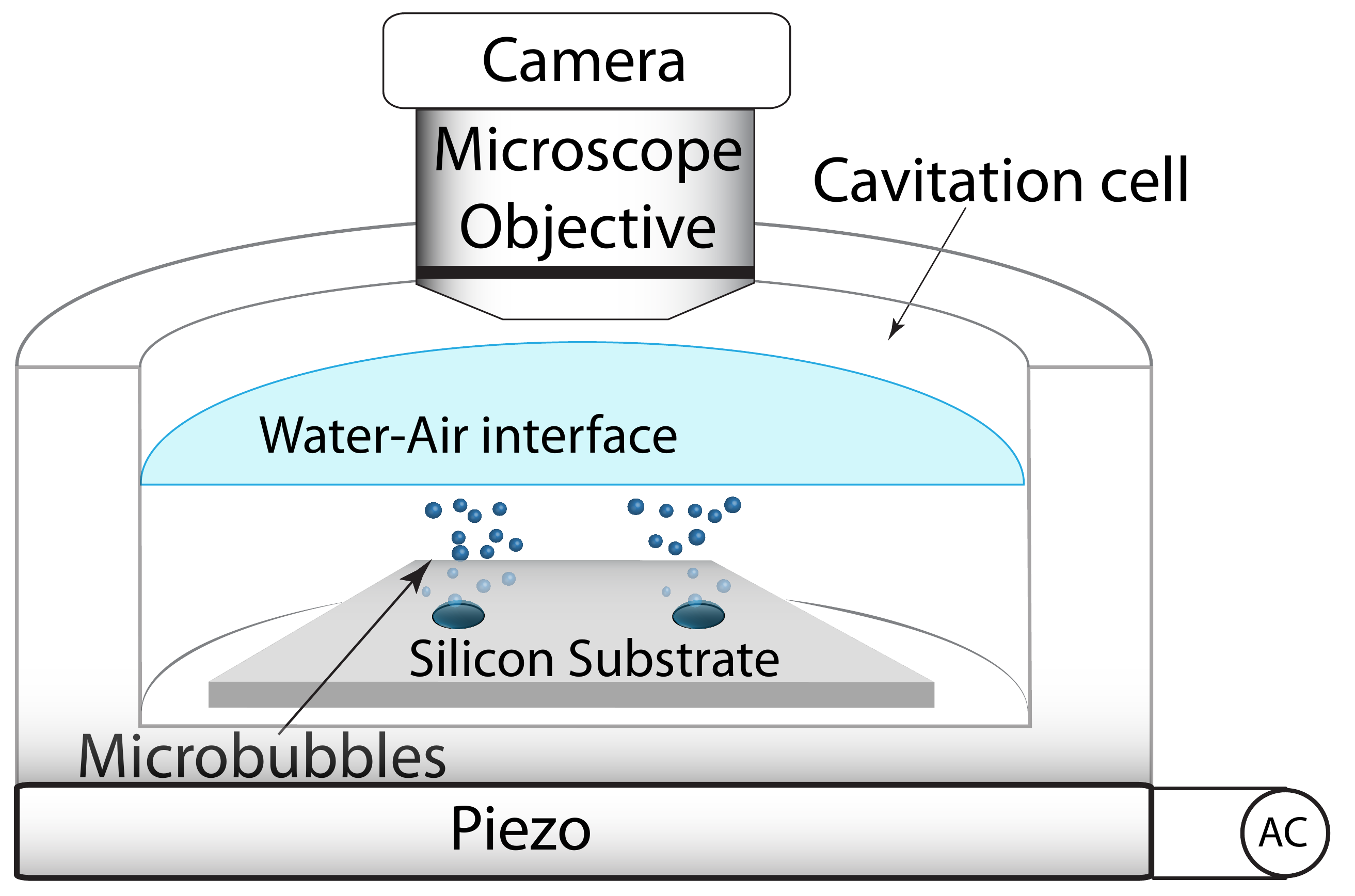}
\caption{Illustration of the experimental setup.}
\end{figure}

The experimental setup presented in Fig.\ref{fig1}, shows the components used to generate cavitation from micropits etched on silicon surfaces. The piezo is controlled by an amplified wave generator signal. This system can provide top and side views of the cavitation events provided by the nucleation of bubbles from artificial crevices etched into a smooth silicon surface.

Chemical and most physical effects of ultrasound arise from the collapse of bubbles in a sonicated liquid. Improvements on sonochemical efficiency in microreactors have been reported recently using pits micromachined on a silicon surface \cite{Fdez5}. 
The main advantage of this method is that the location of cavitation occurrence is known a priori, which allows a detailed study of several relevant aspects of sonochemistry, such as sonoluminescence and other effects such as erosion, cleaning action of bubbles, shock waves, jetting, streaming, etc. \cite{Fdez2,Fdez3,Fdez4}; all of which have been difficult to address in the past, given the rather randomness of cavitating bubbles associated phenomena. 

Links to videos: \href{http://stilton.tnw.utwente.nl/people/alvaro/freezingdrop1.mp4}{50 MB} and \href{http://stilton.tnw.utwente.nl/people/alvaro/freezingdrop1.mp4}{11 MB}.

{\footnotesize
\bibliographystyle{plainnat}

}
\end{document}